\documentclass{elsart}
\usepackage{graphicx} % \usepackage{epsfig}

\usepackage{hyperref}

\newcommand{\be}{\begin{equation}}
\newcommand{\ee}{\end{equation}}
\newcommand{\keyw}[1]{{\bf #1}}

\begin{document}
%\runauthor{aaa}
\begin{frontmatter}
\title{Oxygen diffusion in nanostructured perovskites}

\author[ssc]{P.V. Glyanenko},
\author[ssc]{Yu.M. Kamenetsky},
\author[ssc]{A.P. Nemudry\thanksref{X}},
\author[ssc]{I.L. Zhogin%\corauthref{cor1}
\thanksref{X}},
\author[nano]{H.J.M. Bouwmeester\thanksref{X}},
\author[cat]{Z.R. Ismagilov}

%\corauth[cor1]{Phone: +7(383)3394298}

\thanks[X]{zhogin@inp.nsk.su; nemudry@solid.nsk.su;
h.j.m.bouwmeester@tnw.utwente.nl}

\address[ssc]{Institute of Solid State Chemistry
and Mechanochemistry,
\\ 630128 Novosibirsk, Russia}
\address[nano]{Institute for Nanotechnology, University of Twente,
 The Netherlands}
\address[cat]{Institute of Catalysis, 630090 Novosibirsk, Russia}

\begin{abstract}

Nonstoichiometric perovskite-related oxides (such as ferrites
and cobaltites, etc.) are characterized by fast oxygen transport
at ambient temperatures, which relates to the microstructural
texturing of these materials, consisting wholly of nanoscale
microdomains.

We have developed an inhomogeneous diffusion model to describe
the kinetics of oxygen incorporation into nanostructured oxides.
Nanodomain boundaries are assumed to be the high diffusivity
paths for oxygen transport whereas  diffusion into  the domains
proceeds much slower. Using Laplace transform methods, an exact
solution is found for a ramped stepwise potential, allowing
fitting of the experimental data to theoretical curves
 %for the current vs.\ time.
(in Laplace transforms).
A further model generalization is considered by introducing
additional parameters for the size distribution of domains and
particles.

The model has been applied for qualitative evaluation of oxygen
diffusion parameters from the data on wet electrochemical
oxidation of nano-structured perovskite
SrCo$_{0.5}$Fe$_{0.2}$Ta$_{0.3}$O$_{3-{\rm y}}$ samples.

\end{abstract}
\begin{keyword}
inhomogeneous oxygen diffusion; nanostructured perovskites
\PACS  %\sep
\end{keyword}
\end{frontmatter}

\section{Introduction}
At high temperature grossly nonstoichiometric perovskite related
oxides are solid state solutions with  a high concentration of
structural defects (oxygen vacancies, dopant ions, etc.). As
temperature decreases, structural ordering occurs, producing
either superstructures, in which defects are assimilated as
structural elements, or complex microstructures with a high
density of extended defects  \cite{anderson}.
The formation of the latter microstructures may be described
using the concept of oversaturated solid solutions having a
miscibility gap: oversaturation due to cooling causes unmixing
of the initially homogeneous solid solution. The unmixing can
proceed either by spinodal decomposition or
homogeneous/heterogeneous nucleation \cite{schmalzried}.
Stabilization of nano-sized domains occurs due to a decrease in
mobility of the domain boundaries at $T \leq 0.3\,T_{\rm m}$.
The mechanisms that are responsible for the enhanced thermal
stability against microdomain growth are: (a) solute drag, (b)
grain boundary segregation, (c) pinning by secondary phases and
(d) chemical ordering  \cite{tjong}.
Another approach to the explanation of intrinsic inhomogeneity
of nonstoichiometric and doped oxides related to a strong
tendency toward phase separation in systems with strongly
correlated electrons. Electron conducting oxides with transition
metals having different charges (for example, Mn$^{3/4+}$,
Co$^{3/4+}$, Cu$^{2/3+}$, etc) and magnetic states can break
into a stable state made out of nanoscale coexisting clusters
because of complicated electron-phonon and magnetic interactions
\cite{moreo}.
The Gibbs energy of such a submicro-heterogeneous
(nanostructured) system is essentially influenced by the nature
of the interface between matrix and nucleus. This interface may
be coherent, semi coherent and incoherent. If the interface is
coherent or semicoherent the system behaves as a homogeneous
single phase in terms of diffraction. Microdomain-textured
oxides exemplify such systems rather well \cite{anderson}.
As shown in \cite{alario}, unmixing of nonstoichiometric and
doped oxides with perovskite-related structures produces
microdomains of  5--50 nm in size.

Nanostructured materials may show enhanced ionic conductivity
owing to the high density of interfaces which are enriched by
defects due to the formation of space charge layers \cite{mayer}
or reduction of coordination number of cations in the vicinity
of interface  \cite{wurschum}.

Therefore, nonstoichiometric doped oxides with micro\-do\-main
texture may provide conductive channels for enhanced oxygen
diffusion along the micro\-do\-main boundaries  \cite{bouw,nem}.
Materials with high oxygen mobility are of interest for
applications such as oxygen electrolytes, electrodes, sensors
and membranes for oxygen separation  \cite{bouw}.

Numerous experimental proofs  of enhanced diffusion along grain
boundaries (including radiography, \cite{fens,acht,hofm}) have
initiated theoretical investigations of inhomogeneous diffusion
in polycrystalline materials  \cite{Fisher,boks,gegu}.
The first mathematical description of fast diffusion along grain
boundaries accompanied by penetration into the grains by means
of bulk diffusion (without phase transition though) is shown in
\cite{Fisher,Whipple}.
%%%%%
In \cite{nem_gold,nem_schoell}, a mathematical model of the
two-phase oxidation of MDT nonstoichiometric perovskite has been
considered. The crystal is conceived as an assembly of parallel
domain boundaries, which are considered to be high diffusivity
paths. Oxidation proceeds as a result of rapid oxygen transport
along the domain boundaries and two-phase oxidation of the
microdomains.

Unlike the previous papers, here we assume that oxidation of
microdomains occurs as a result of one phase reaction.
%In this paper,
We employ a model for oxygen transport in the MDT perovskites in
the course of single phase oxidation, assuming that diffusion
along the nanodomain boundaries is much faster than diffusion
 into the domains.
Since nanostructured perovskites exhibit fast oxygen transport
even at room temperature, the model was applied  for the
evaluation of oxygen diffusion parameters from data of wet
electrochemical oxidation of nano-structured perovskite
SrCo$_{0.5}$Fe$_{0.2}$Ta$_{0.3}$O$_{3-{\rm y}}$ samples in 1M
KOH at ambient temperatures.

Potentiostatic method \cite{gur,wen} is considered to be a
%universal and
convenient tool for studying oxygen transport in perovskites
(where electron conductivity far exceeds the ion one). In the
present study we have used a version of this method.

\section{Experimental}

Kinetic studies were carried out for
SrCo$_{0.5}$Fe$_{0.2}$Ta$_{0.3}$O$_{3-\mathrm{y}}$
nonstoichiometric perovskite. The samples were synthesized by
solid state reaction from the corresponding metal oxides and
carbonates with preliminary homogenization of the starting
materials in a planetary ball mill. A stoichiometric mixture of
the powders was calcined at  900~$^\circ$C, pressed in pellets,
annealed in air at 1400~$^\circ$C for 6 hours and cooled in the
furnace.

For kinetic studies, the sintered sample
%SrCo$_{0.5}$Fe$_{0.2}$Ta$_{0.3}$O$_{2.7}$
was annealed at 950 $^\circ$C in a quartz ampoule in dynamic
vacuum ($P \sim 10^3$ Pa). After that, the quartz ampoule was
closed and placed in water for rapid quenching of the sample to
room temperature.

Experiments were performed at room temperature (25 $^\circ$C) in
potentiostatic mode (three electrode cell, 1M KOH electrolyte,
Hg/HgO reference electrode) with working electrodes of
polycrystalline material (17--18 mg) pressed into Pt grids along
with 1 wt.\% of Teflon and 15--20 wt.\% of acetylene black. The
working electrode was placed in a cell at the set temperature
and was maintained up to an equilibrium potential   $U_0$. At
$t=0$ a voltage pulse between the working electrode and
reference electrode has been applied. Re-equilibration through
diffusion of oxygen into the working electrode material takes
place. The rate of re-equilibration can be measured easily by
monitoring the electric current passing through the cell, see
Fig.~1.
%$J_\mathrm{exp}(t)$. %, as a function of time.

%%%%%%%%%%%%%%%%%%%%%%%%%%%%%%%%%%%%%%%%%%%%%%%%%%%%%%%%%%
\begin{figure}
  \centering
  \includegraphics*[width=100mm]{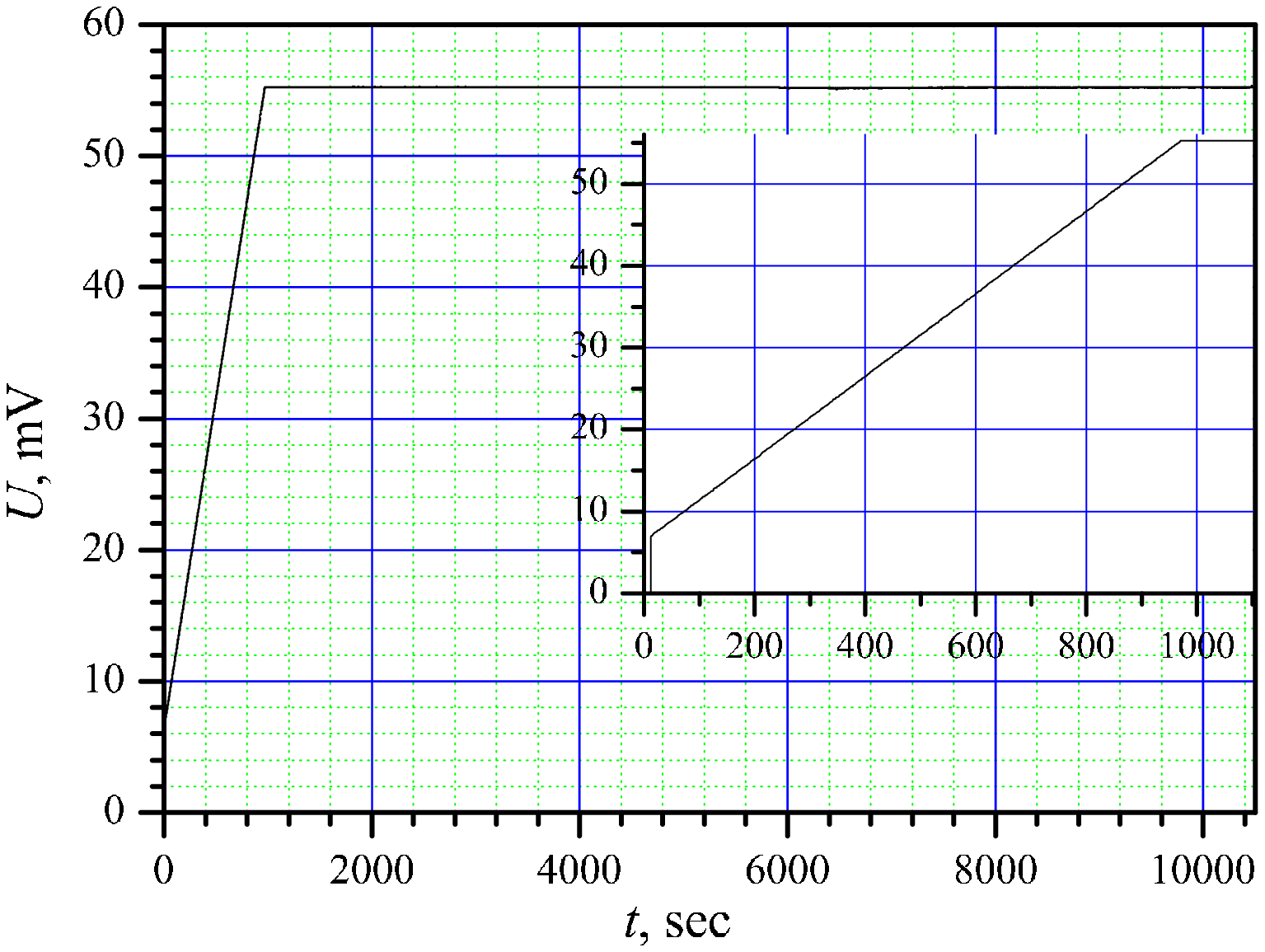}\\
  \includegraphics*[width=100mm]{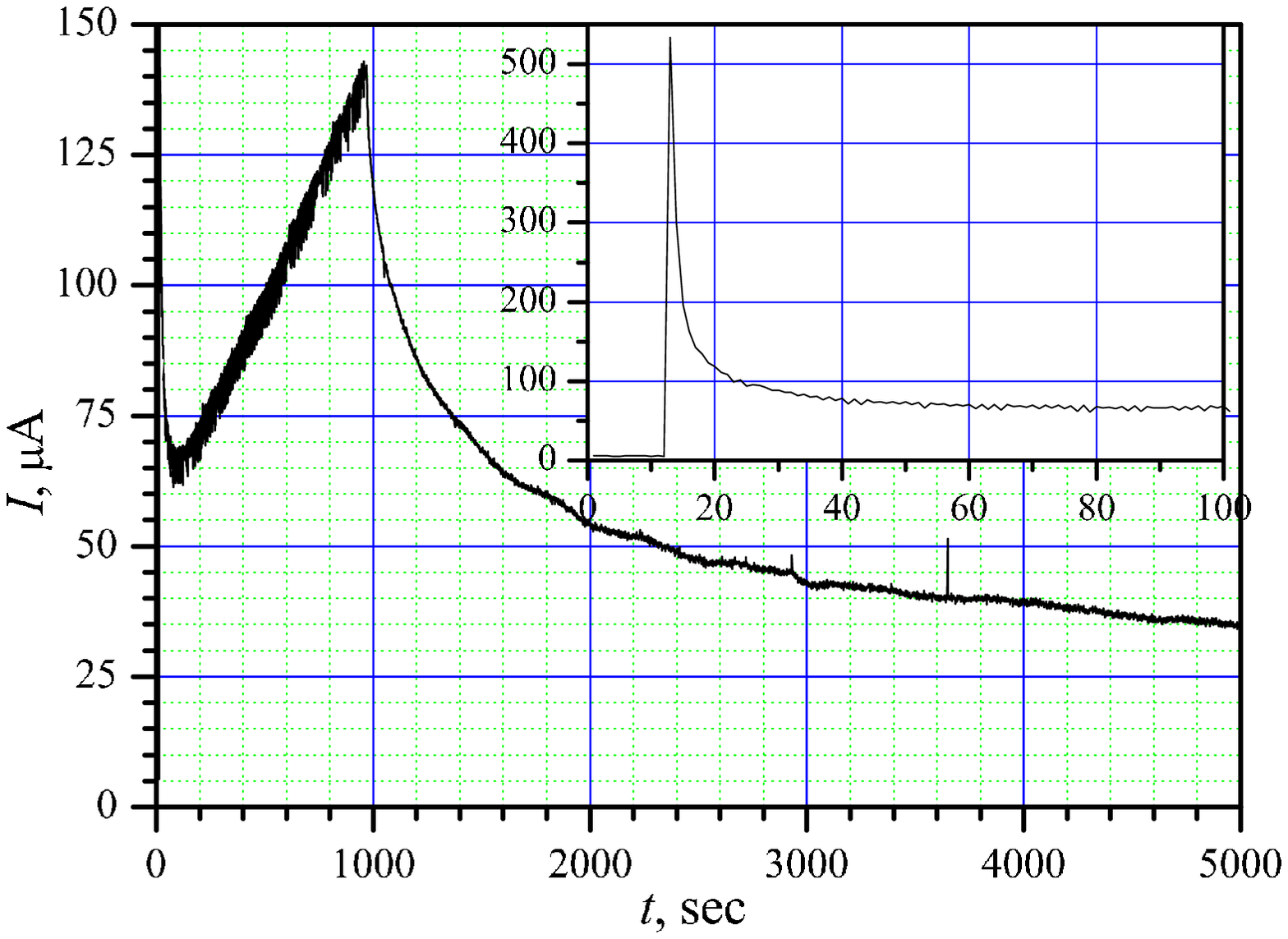}\\
  \caption{Applied potential (a) and current (b) versus time. }
\end{figure}
%%%%%%%%%%%%%%%%%%%%%%%%%%%%%%%%%%%%%%%%%%%%%%%%%%%%%%%%%%

\section{Theoretical model}
Similar to the model of spheres filling a half-space (grains in
a thick metal plate, see  \cite{boks}), we consider a
 %`matreshka'
 model, where small spheres are inside  large ones
(nano-domains inside the perovskite powder particles). The
simple geometry of our model allows us to derive an exact
solution for the current passing through the cell using the
Laplace transform method.

Inverse Laplace transformation to determine the original
function is not possible in analytical form (in analytical
functions; this is the reason why one usually analyzes only the
asymptotic current behavior at short and long times). Though
there exist computational methods to do reverse Laplace
transformations \cite{huep}, they require intensive calculations
(double Fourier transforms). Therefore, we have chosen another
way of model comparison with experiment. After we perform
numerical Laplace transformation of measured current function,
$J_\mathrm{exp}(t)$, (as well as potential function, if it is
measured and is not given ``analytically'' with potentiostate)
we adjust transform, $\hat{J}_\mathrm{exp}(p)$, with model
curves $\hat{J}_\mathrm{th}(p)$ (which depend on the applied
potential profile).

However, computer inversion of Laplace transform may be useful
for verifying the best approximation,
 or for the better choosing among several minima, if they occur.

A two-level model of spheres is used to describe the current
passing through an electrochemical cell under potential of a
chosen profile.

\subsection{Diffusion in a spherical particle}
 The diffusion equation for a spherical particle, when the space
derivatives (Laplacian) are reduced to differentiation over the
radius, is written as follows:
 \be \label{sph}
 \frac{\partial(x\, c)}{\partial t} =
 D \frac{\partial^2(x\, c)}{\partial x^2 } \, .
  \ee
 Here  $c(t,x)$ is the concentration of the diffusing species,
 % component,
  $R$  the particle radius, and $x$
 the radial coordinate,  $0\leq x\leq R$. For initial and
boundary conditions we have
\[
 c(t<0,x)=0\,, \ c(t,x=R) = c_R(t) \, .
 \]

 Using the Laplace transform of  $c(t)$
 \[
\hat{c}(p,x) \equiv \int^\infty_0 c(t,x) e^{-pt} \,\mathrm{d} t
 \]
turns partial derivatives  (\ref{sph}) into an ordinary
differential equation
 \be \label{sphl}
   \frac{\partial^2(x \hat{c})}{\partial x^2 }
  =\frac{p}{D}x \hat{c} .
 \ee

 Applying boundary condition,
 $\hat{c}(p,R)= \hat{c}_R(p)$,
one can easily find the exact solution of equation (\ref{sphl}),
  \be \label{cpx}
 \hat{c}(p,x)= \hat{c}_{R}(p)\,
 \frac{\kappa R  }{\sinh(\kappa R)}\,
 \frac{\sinh(\kappa x)}{\kappa x} \, , \
 \ \kappa= \sqrt{p/D} \, .
 \ee

Diffusion current (and electric one, if diffusing component
carriers charge $q$),  passing through the whole particle
boundary, may be expressed as follows:
\[
 J(t) = 4\pi q \int_0^R \frac{\partial c(t,x)}{\partial t}
 x^2 \,\mathrm{d} x \, , \]
\be
\label{jla}
\hat{J}(p) =4 \pi q \,
\frac{p \hat{c}_R(p)\, R}{\sinh(\kappa R)}
\int_0^R \sinh(\kappa x) \,x \,\mathrm{d} x \, .
\ee

One may expect that the diffusion coefficient is independent of
concentration, provided that the oxygen content in the
perovskite lattice changes only slightly,  e.g., $\Delta
\mathrm{y} < 0.1$. For this purpose the amplitude of potential
pulse, $U_A$, applied to the cell, should be small enough (this
is determined by the ratio of sample mass and charge). We assume
that the access of oxygen ions to the particles is not limited
and, hence, there is a linear dependence between concentration
and the potential
\[
c_{R}(t) \propto U(t) \mbox{ or }
c_{R}(t)/c_A = U(t)/U_A \, .
\]
In experiments we used potential steps of the following form
(see Fig.~1a):
%%%%%%%%%%%%%%%%%%%%%%%
\[ U(t)/U_A = H(t;h,t_0) \equiv \left\{
\begin{array}{cl}
0, & t<0 \\
h + (1-h)t/t_0, & 0\leq t  \leq t_0\\
1, & t > t_0
\end{array} \right. ;
\]
this gives the Laplace transform ($0\leq h \leq 1$):
 \be \label{crl} p \,
c_R(p)/c_A=h + (1-h)(1-\e ^{-p t_0})/(p t_0)
 \ \ (\,
\stackrel{p\to 0} {\longrightarrow} 1)\, .
 \ee
 Introducing value  $Q\equiv
\hat{J}(0)=4 \pi R^3q c_A/3$ for a total charge passed through
the cell, and using equations  (\ref{cpx}), (\ref{jla}) and
(\ref{crl}) we obtain
 ($  \tau \equiv R^2\!/D $
 is a characteristic diffusion time
for a  sphere of radius
 $R$; $\sqrt{p\tau} = \kappa R$)
  \be
\label{jlb} \hat{J}(p)/Q =
 %\left(h + (1-h)\frac{1-\e ^{-pt_0}}{p t_0}\right)
 p\, c_R(p)/c_A \;
 F(\sqrt{p\tau})\, ,\ee
  \be
  \mbox{where \ }
\label{fdef}
 F(y) \equiv
\frac{3 \coth(y)}{y} - \frac{3}{y^2} \ \ (\, \stackrel{y\to 0}
{\longrightarrow} 1-y^2/15+\cdots)\, \, .
 \ee

\subsection{Inhomogeneous diffusion model: spheres of two types}
Let us imagine that particles in the sample are non-uniform (see
Fig.~2) and contain domains of a typical size $r$, where
diffusion rate is small, $D_1$, and well as regions neighboring
with domains boundaries, where diffusion rate is high, $D_2$.
More exactly,  let us consider $D_2$ as an effective diffusion
coefficient for a porous particle, obtained, when all domains
are removed, as if they are non-permeable for diffusion, i.e.
 $D_1= 0$.
 This effective coefficient may change not only with
temperature, but also with parameters related to the nano-domain
structure, such as oxygen content, quenching rate, etc.

%%%%%%%%%%%%%%%%%%%%%%%%%%%%%%%%%%%%%%%%%%%%%%%%%%%%%%%%%%
\begin{figure}
  \centering
  \includegraphics*[width=100mm]{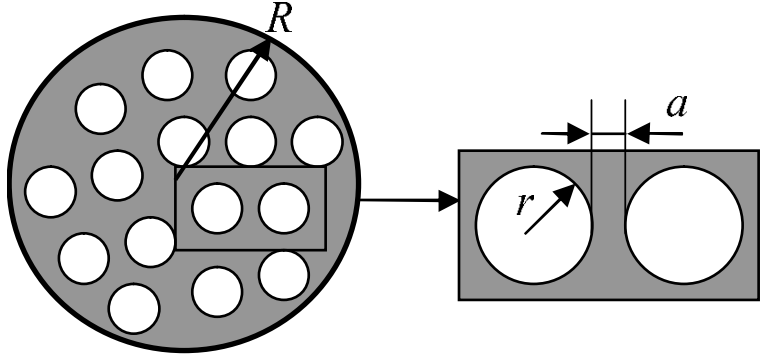}\\
  \caption{Regions of slow (white) and fast (grey) diffusion}
\end{figure}
%%%%%%%%%%%%%%%%%%%%%%%%%%%%%%%%%%%%%%%%%%%%%%%%%%%%%%%%%%

If diffusion coefficient $D_1$ is small but not equal to zero,
then diffusion equation should reflect oxygen incorporation
inside the domains. One may assume that the concentration along
the boundary of each domain is approximately the same (since
transport to the boundary is fast) and is equal (or at least
proportional) to the current local concentration:
\[
 c_{1r}(t) = c_2(x,t) \, . \]
Therefore,
%regarding equations  (\ref{jla}), (\ref{jlb}) and(\ref{fdef}),
%
 we may write  effective diffusion equation in a
heterogeneous particle instead of equation (\ref{sph}):
 \[
v_2\left( \frac{\partial c_2}{\partial t} -
 D_2 \frac{\partial^2(x c_2)}{x \,\partial x^2 }
\right) =
  - n_{\mathrm{d}} J_1(t,x)/q
  \, .
  \]
 Here $n_{\rm{d}}$ is the concentration of domains, $J_1/q $
  the diffusion current into a domain,
 $v_1=n_{\rm{d}}\, 4\pi r^3/3, \ v_2=1-v_1$
are  volume fractions for %domains and  boundaries.
regions of slow and fast diffusion, respectively. After Laplace
transformation we obtain (taking into account equations
(\ref{crl}), (\ref{jlb})):
 \be \label{sphm}
   \frac{\partial^2(x \hat{c}_2)}{\partial x^2 }
  =\frac{p}{D_2}x \hat{c}_2\left(1 + \alpha F(\sqrt{p\tau_1})
  \right)\, ,
 \ee
 where $\alpha = v_1/v_2$ %\sim r/3a$
is the ratio of above mentioned volume fractions (see Fig.~2).

As it is mentioned in \cite{boks}, equilibrium concentrations
(of diffusant) along the boundaries and inside the domains (i.e.
grains) may not coincide. Therefore, dimensionless parameter
$\alpha$ may include not only geometry but also the
concentration factor.

One may derive the final expression for the current using
equation (\ref{sphm}) (it is necessary to remember that the
particle is inhomogeneous):
 \begin{eqnarray}
 \nonumber
  \hat{J}(p) &=& 4\pi R^2\, D_2
 \left. \frac{\partial \hat{c}_2}{\partial x} \right|_{x=R}\\
 \label{jlc}    &=& Q\,
\frac{p \hat{c}_{2R}(p)}{c_{A}} \,
\frac{1+\alpha f(p\tau_1)}{1 + \alpha} \,
f(p\tau_2[1+\alpha f(p\tau_1)])
\, .
 \end{eqnarray}
Here notations
 $f(y) \equiv F(\sqrt{y}), \
 \tau_1 \equiv r^2\!/D_1, \ \tau_2\equiv
 R^2\!/D_2$
 (see equations
 (\ref{crl})--(\ref{fdef}))
 are introduced.

%%%%%%%%%%%%%%%%%%%%%%%%%%%%%%%%%%%%%%%
\subsection{Assembly of particles, particle size distribution}
If powder particles are of different sizes, then the current
expression (\ref{jlb}) and expression  (\ref{jlc})
 should be integrated with the function of volume (mass)
 distribution of particles,   $M(R)$
(norm per unit):
 \[
\hat{J}(p)/Q =  \frac{p\, \hat{c}_{R}(p)}{c_{A}} \,
\int_0^\infty F(\kappa R) \, M(R) \,\mathrm{d} R\, .
 \]
If the `relative'  dispersion of the distribution,
 \[ \Delta_R=\langle(R - R_0)^2\rangle/R^2_0
 \ \ (R_0 =\langle R \rangle )\, , \]
 is small, we leave the first
correction term, changing functions in (\ref{jlc}) or
(\ref{jlb})  as follows (prime means differentiation by
argument):
 \be \label{jld} F(\kappa R) \to
F^*(\kappa R_0)= F(\kappa R_0) + (\kappa R_0)^2 F''(\kappa R_0)
\Delta_R/2 \, , \ee
 \[ \mbox{where \ }
 y^2 F''(y) =
6(1 +y^2/\sinh^2y)(1 + y/\tanh y) -24/y^2 \, .
 \]
One may obtain this correction by e.g.\ integration with formal
distribution (delta functions):
\[
M(R) = \delta(R-R_0) + \delta''\!(R-R_0)\, \Delta_R \, R_0^2 /2
\, ,
 \]
which has proper first two moments. Saddle-point integration
with the Gaussian distribution
 $M(R) \propto
 \exp(-(R/R_0-1)^2/2\Delta_R)$
 gives the same result.

In the similar manner one may correct for the size dispersion of
nano-domains, $\Delta_r$.

%%%%%%%%%%%%%%%%%%%%%

\section{Experimental results and computation}
% (see Fig.~1).
%%%%%%%%%%%%%%%%%%%%%%%%%%%%%%%%%%%%%%%%%%%%%%%%%%%%%%%%%%
\begin{figure}
  \centering
  \includegraphics*[width=100mm]{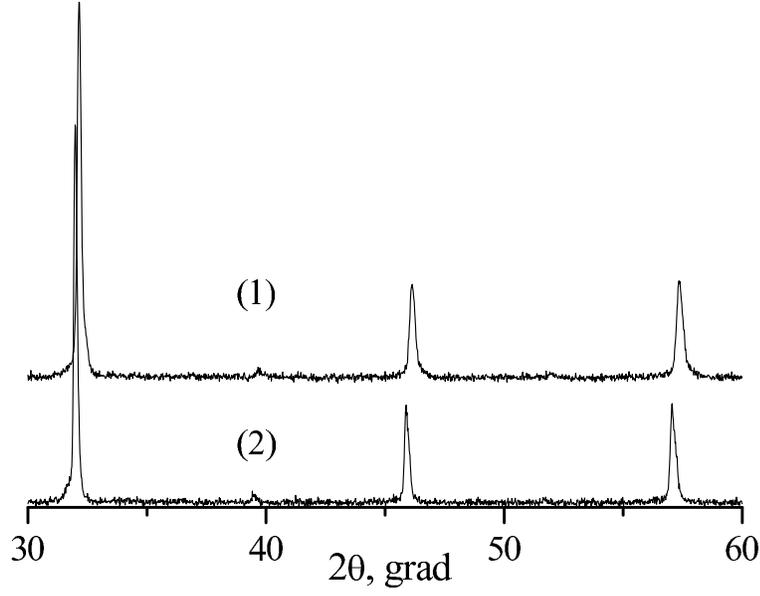}\\
  \caption{Diffraction patterns of samples
 SrCo$_{0.5}$Fe$_{0.2}$Ta$_{0.3}$O$_{3-\mathrm{y}}$:
  (1) as sintered,  (2) after annealing at
  950~$^\circ$C and quenching in vacuum ($P\sim10^3$ Pa).  }
\end{figure}
%%%%%%%%%%%%%%%%%%%%%%%%%%%%%%%%%%%%%%%%%%%%%%%%%%%%%%%%%%

According to the X-ray diffraction (Fig.~3) and iodometric
titration data, quenched samples of
SrCo$_{0.5}$Fe$_{0.2}$Ta$_{0.3}$O$_{3-\mathrm{y}}$
had the expanded cubic unit cell parameters (from 3.934 \AA\ to
3.952 \AA) and reduced oxygen content (from 2.92 to 2.70) with
respect to slowly cooled ones.

XRD suggests single phase behaviour, while the HREM data shows
that the samples possess a micro-domain texture (Fig.~4) with a
typical domain size of about 10 nm.
%%%%%%%%%%%%%%%%%%%%%%%%%%%%%%%%%%%%%%%%%%%%%%%%%%%%%%%%%%
\begin{figure}
  \centering
  \includegraphics*[width=100mm]{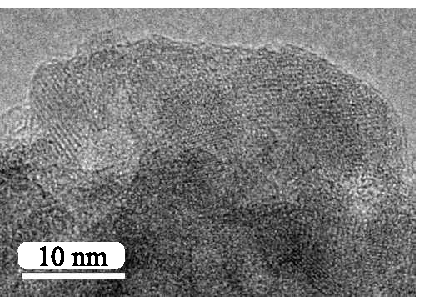}\\
  \caption{High resolution image
  of SrCo$_{0.5}$Fe$_{0.2}$Ta$_{0.3}$O$_{2.7}$
  sample possessing micro\-do\-main texture.}.
\end{figure}
%%%%%%%%%%%%%%%%%%%%%%%%%%%%%%%%%%%%%%%%%%%%%%%%%%%%%%%%%%

The mechanism of electrochemical oxidation of
SrCo$_{0.5}$Fe$_{0.2}$Ta$_{0.3}$O$_{3-\mathrm{y}}$
%%%%%
has been studied with chronopotentiometry combined with {\em in
situ} X-ray diffraction.
The results of evaluation of the {\em in situ} data are shown in
Fig.~5. Monotonous changes in the unit cell parameters and
potential vs.\  inserted oxygen (x) (${\rm x}=n/2$, where $n$ is
a charge transfer) provide evidence of one phase mechanism of
oxidation \cite{glya}.
%%%%%%%%%%%%%%%%%%%%%%%%%%%%%%%%%%%%%%%%%%%%%%%%%%%%%%%%%%
\begin{figure}
  \centering
  \includegraphics*[width=100mm]{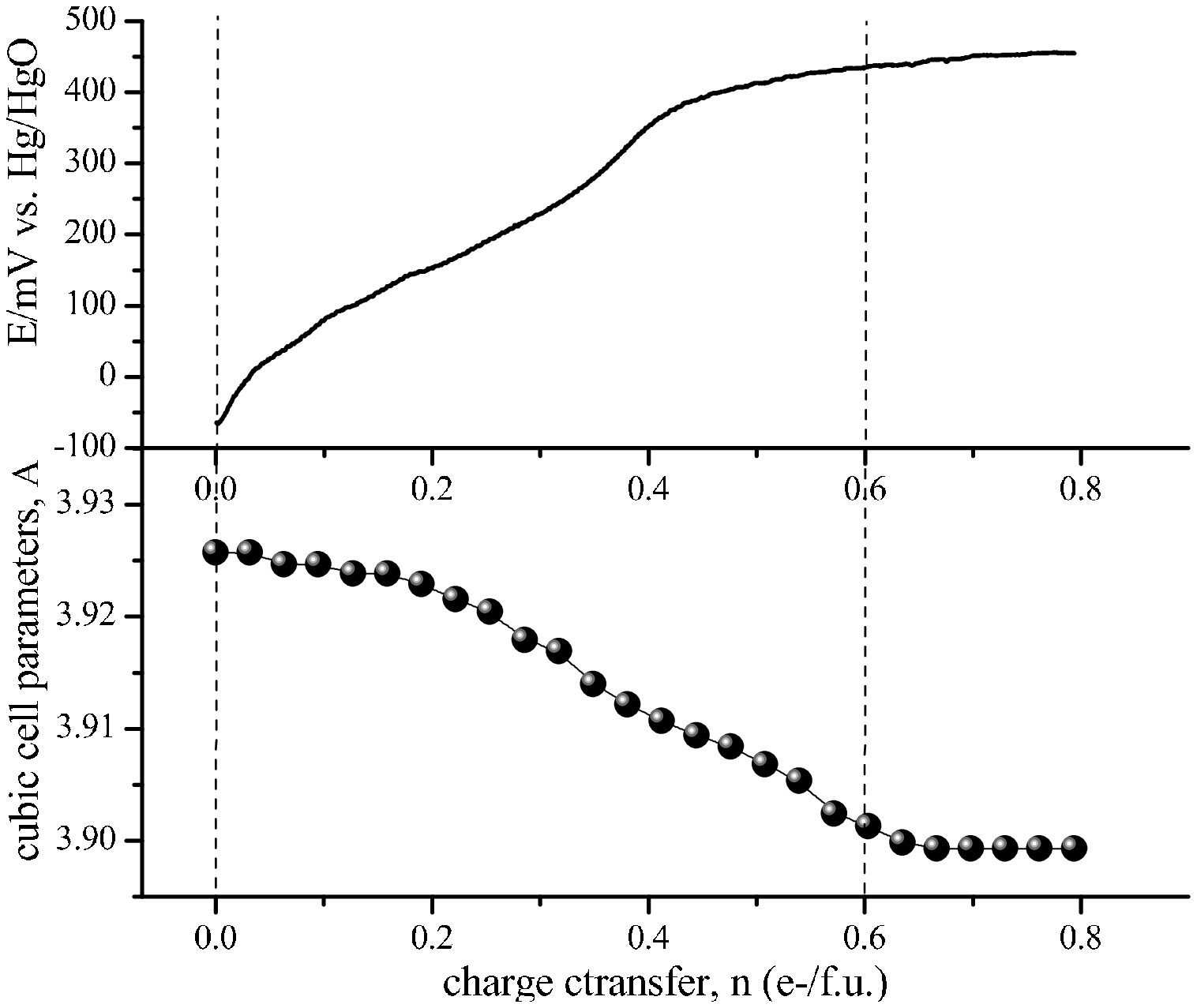}\\
\caption{In situ anodic oxidation of
 SrCo$_{0.5}$Fe$_{0.2}$Ta$_{0.3}$O$_{2.7+\mathrm{x}}$:
 %SrCo0.5Fe0.2Ta0.3O2.7+x:
(a) potential $E$ vs.\ charge transfer $n$, (b) change of
unit-cell parameters with $n$, $n/2=\mathrm{x} $.}
\end{figure}
%%%%%%%%%%%%%%%%%%%%%%%%%%%%%%%%%%%%%%%%%%%%%%%%%%%%%%%%%%

For kinetic studies, the samples were powdered, and particle
size distribution was measured (see Fig.~6).
%%%%%%%%%%%%%%%%%%%%%%%%%%%%%%%%%%%%%%%%%%%%%%%%%%%%%%%%%%
\begin{figure}
  \centering
  \includegraphics*[width=100mm]{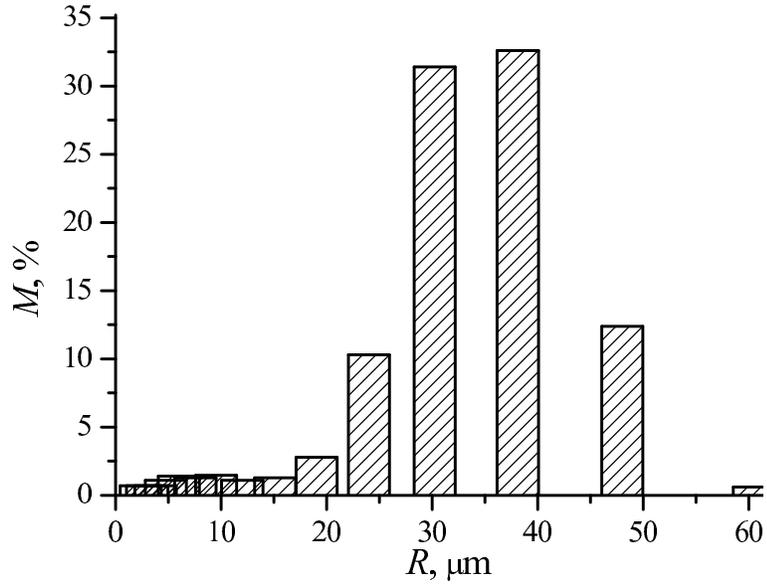}\\
  \caption{Mass distribution of particles over size. }
\end{figure}
%%%%%%%%%%%%%%%%%%%%%%%%%%%%%%%%%%%%%%%%%%%%%%%%%%%%%%%%%%

Fig.~1 shows $U(t)$ and $J_\mathrm{exp}(t)$ functions {\em
versus} time, obtained in one of experiments. The time of
current measurement averaging was 1 second. This obstacle
somewhat distorted the current plot at short times.

Computations were performed in MatLab. For the data, presented
in Fig.~1, potential step parameters are the following (see
equations (\ref{crl}) and (\ref{jlc})):

  \quad h = 7/55.25;  t\_0 = 958;  \%\% sec

After $\hat{J}_\mathrm{exp}(p)/Q$ was calculated (exponential
slope at long times ($t>$3000 s) was considered analytically),
correction for the time of micro ammeter integration (1 s) was
introduced:

  \quad Jep = Jep.*p./(1 - {\bf exp}(-p));

Then optimization over three parameters entering the model
function of current (equation (\ref{jlc})) was done:
\begin{tabbing}
 \quad \= \quad \= \quad \= \quad   \kill
  \> par0 = [tau1,tau2,alpha];\\
  \> options = optimset('TolFun',1e-10,'TolX',1e-6, ...\\
 \>\>\> 'MaxFunEvals',1e3,'MaxIter',1e3);\\
 \>[par1,fval,exitflag,output] = ...\\
 \>\> \keyw{fminsearch}%%
('\keyw{diff\_fun}',par0,options,p,Jep,handle);\\
\> $\cdots$\\
\> \keyw{function}
ff = \keyw{diff\_fun}(param,p,Jep,handle)\\
\> $\cdots$ \\
\> ff = \keyw{sum}(\keyw{abs}(Jtp - Jep));
\end{tabbing}
Computing results are shown in Fig.~7, model parameters for
theoretical curve (dotted line) being as follows:
 \[
 \tau_1=49 \ \rm{s}, \
 \tau_2=19\cdot 10^3 \ \rm{s}, \
 \alpha= 4.8 \, .\]
Using values
 $R=32$ $\mu$m, $r \sim 10 $ nm,
 one may estimate diffusion coefficients
 \[
 D_2 = 5\cdot 10^{-10} \ \rm{cm}^2/\rm{s} ,
  \ D_1= 2\cdot 10^{-13} \ \rm{cm}^2/\rm{s}.
 \]
%%%%%%%%%%%%%%%%%%%%%%%%%%%%%%%%%%%%%%%%%%%%%%%%%%%%%%%%%%
\begin{figure}
  \centering
  \includegraphics*[width=100mm]{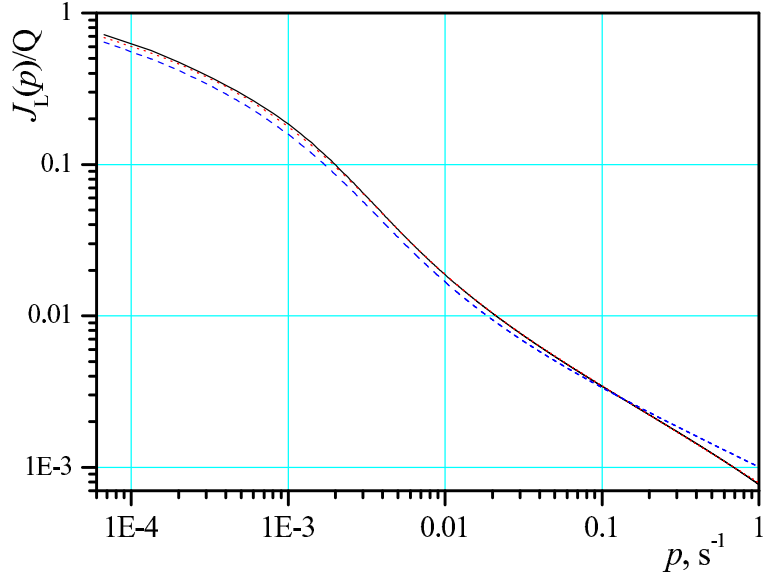}\\
 \caption{Normalized Laplace transform
 % $\tau_1, \tau_2, \alpha$;
 of experimental data (solid; $Q=0.44$ C) and fitting model curves
  for inhomogeneous (dot) and homogeneous
  (dash; $\alpha = 0$) model.}
\end{figure}
%%%%%%%%%%%%%%%%%%%%%%%%%%%%%%%%%%%%%%%%%%%%%%%%%%%%%%%%%%

Let us note that fitting by the homogeneous model gives
essentially larger errors (see dashed curve in Fig.~7), and
yields another value:
 $\tau_2=15\cdot 10^4 $~s.

%%%%%%%%%%%%%%%%%%%%%%%%%%%%%%%%%%%%%%%

\section{Discussion and conclusions}
%Model (see Fig.\ 2) and model image function (see equation
%(\ref{jlc})) both have recurrent (matreshka-like) structure:
%small spheres inside the large one, and the function ($F$)
%inside of the function. The model developed
% has been applied for wet electrochemical oxidation of
%nanostructured perovskite, which is promising as oxygen
%permeable membrane materials. Oxygen diffusion parameters were
%evaluated regarding the current-time measurements.
%
%We believe that using voltage pulses $U(t)$ of different profile
%(shape) one may increase the accuracy of determining the
%numerous fitting model parameters (finite pulses give $Q = 0$,
%but normalization is possible at maximum  $\hat{J}(p)$).
%
%Linear potential growth with no initial step is preferable,
%since there is no initial current singularity,
% $ J(t) \sim 1/\sqrt{t}$.
%However, we have found that potential growth consists of small
%steps. This is the typical feature characterizing our measuring
%devices -- feed back potentiostate and its programmer. These
%steps cause noticeable fluctuations on the current curve.
%
%In future we are going to apply potential impulse through the
%RC-chain using large capacities.
%
%
%We believe that our experimental and theoretical approach allows
%one to determine various
% parameters of oxygen diffusion in
%nanostructured perovskite samples using relatively simple wet
%electrochemical setup.

As we have already noted, nonstoichiometric and/or doped oxides
are high temperature systems, since solid solutions are stable
at relatively high temperatures only. At lower temperatures the
high free energy related to the high concentration of defects
(oxygen vacancies, guest ions) can be reduced either by defects
ordering and their localization as structural elements
(superstructures formation), or by their elimination or
precipitation (nanostructuring, microdomain texturing)
\cite{anderson}. The difference between the MDT phases and
ordered superstructures is that in MDT oxides part of disorder
is already accumulated at low temperatures as excess interfacial
energy (disordered lattice in the vicinity of interfaces).
Therefore, the lower is working temperature, the more essential
is the MDT oxide advantage with regard to oxygen transport. That
is why MDT oxides are promising materials for low and moderate
temperature oxidative catalysis, oxygen permeable membranes for
partial oxidation of hydrocarbons, SOFC's electrodes, sensors
etc.

With our developed model, we have shown that nanostructuring
allows inhomogeneous oxygen diffusion in domains and along the
interfaces; the difference in diffusion coefficients may attain
several orders of magnitude. The most important issue is that
stationary flux of oxygen ions through a MDT perovskite
membranes at temperatures lower than the point of order-disorder
transition is determined mostly by coefficient $D_2$.

Since activation energy $E_{\rm{a}}$, necessary for the oxygen
ions migration along the grain boundaries (g.b.), may be
essentially lower than that for the bulk (b.) diffusion,
  $E_{\rm{a}} ({\rm g.b.}) \sim 1/2\, E_{\rm{a}}({\rm b.})$
\cite{Kofstad},
 oxygen permeable
membranes made of MDT oxides are able to provide several orders
of magnitude higher oxygen fluxes at the working temperatures
below the order-disorder transition point. Let us mention though
that nanostructuring is not the sufficient condition for
intensive oxygen transport along the interfaces. For example,
alkali/rare earth manganites possess a well-developed
microstructure, but oxygen mobility in them is very low, whereas
alkali/rare earth ferrites, and in particular cobaltites show
unusually high oxygen transport [8]. Apparently, the transport
properties of oxides are influenced by the interfacial energy,
which is dependent on a set of parameters (M--O bond strength,
coordination of cations, their charges and size, electronic
configuration, magnetic state of cations, etc.). In case if
interfacial energy is high enough (disordered interface), one
may expect high permeability of oxygen. If interface is ordered
(low interfacial energy), most likely it will be a barrier for
the ion diffusion.

We shall further study the kinetics of MDT perovskites oxidation
at various temperatures to determine the activation energy for
oxygen ions migration along the interfaces and inside the
domains and develop the method for measuring oxygen diffusion
parameters in the nanostructured materials. Comparing the data
obtained at low temperature by means of relatively simple wet
electrochemical technique with the data related to oxygen
permeability at high temperatures, we might define the mechanism
of oxygen transport in MDT oxides, elucidate the factors
determining the high values of oxygen fluxes in these materials
and might open strategies to develop new oxygen-conducting
materials operating at moderate temperature.


\begin{thebibliography}{555}
\bibitem{anderson} J.S. Anderson, The thermodynamics and theory
 of nonstoichiometric compounds,
  in:  A. Rabenau, (Ed.), {\em Problems of Nonstoichiometry},
  North-Holland Publ. Co, Amsterdam, 1970, p.1.

\bibitem{schmalzried} H. Schmalzried, in: H.F. Ebel (Ed.),
{\em Solid State Reactions}, Verlag Chemie, 1981, p.167.

\bibitem{tjong}
S.C. Tjong, Haydn Chen, {\em Materials Science and
Engineering\/} {\bf R 45} (2004) 1--88.

\bibitem{moreo}
E. Dagotto, T. Hotta, A. Moreo, {\em Physics Reports\/}
 {\bf 344} (2001) 1--153.

\bibitem{alario}
M.\'A. Alario-Franco, J.M. Gonzalez-Calbet, M. Vallet-Regi,
J.-C. Grenier, {\em J. Solid State Chem.} {\bf 49} (1983) 219.



%\bibitem{heitjans}P. Heitjans, S. Indris, {\em J. Phys.: Condens.
%Matter\/} {\bf 15} (2003) R1257.

\bibitem{mayer}
J. Maier, {\em Prog. Solid State Chem.\/} {\bf 23} (1995)
171-263.

%\bibitem{gleiter}

\bibitem{wurschum}
R. W\"urschum,  {\em Rev. Metall.\/} {\bf 96} (1999) 1547.

\bibitem{bouw}
H.J.M. Bouwmeester, A.J. Burggraaf, Dense ceramic membranes
for oxygen separation,
in: A.J. Burggraaf and L. Cot (Eds.),
 {\em Fundamentals of Inorganic Membrane Science and Technology},
Elsevier, Amsterdam, 1996, p. 435.

\bibitem{nem}
A. Nemudry, E.L. Goldberg, M. Aguirre and M.A. Alario-Franco,
{\em Solid State Sciences\/} {\bf 4} (2002) 677.

\bibitem{fens} P. Fenshom, {\em Australian Just. Sc. Research\/}
 {\bf 3}  (1950) 105.

\bibitem{acht} M. K. Achter and R. Smoluchowski,
{\em  Phys. Rev.\/} {\bf 76}  (1949) 470.

\bibitem{hofm} R. E. Hofmann and D. Turnbull,
  Lattice and grain-boundary self-diffusion in silver,
 {\em Journ. of Appl. Phys.\/} {\bf 22}
No.\ 5 (1951) 634.

\bibitem{Fisher}
J.C. Fisher, {\em J. Appl. Phys.\/} {\bf 22} (1951) 74.

\bibitem{boks}
B.S. Bokshtein, A.I. Magidson and I.L. Svetlov, {\em Fizika
Metallov i Me\-tal\-lo\-vedenie\/} {\bf 6} (1958) 1040. In
Russian.

\bibitem{gegu}
Ya. E. Gegusin. {\em Macroscopic defects in metalls} (Moscow,
1962). In Russian.


\bibitem{Whipple}
R.T.P. Whipple, {\em Phil. Mag.\/} {\bf 45} (1954) 1225.

\bibitem{nem_gold}
E. Goldberg, A. Nemudry, V. Boldyrev, R. Sch\"ollhorn, {\em
Solid State Ionics\/} {\bf 110} (1998) 223.

\bibitem{nem_schoell}
E. Goldberg, A. Nemudry, V. Boldyrev, R. Schollhorn, {\em Solid
State Ionics\/} {\bf 122} (1999) 17.

\bibitem{gur}
S. Sunde, K. Ni\c{s}ancio\v{g}lu, T. M. G\"ur,
{\em J. Electrochem. Soc.\/} {\bf 143} No.\ 11 (1996) 3497.

\bibitem{wen}
C. J. Wen, C. Ho, B. A. Boukamp, I. D. Raistick, W. Weppner, and
R. A. Huggins, {\em Int. Metals Rev.\/} No. 5 (1981) 253.

 \bibitem{huep} B. H{\"u}pper and E. Pollak,
  A new method for numerical inversion of the Laplace transform,
 \href{http://www.arXiv.org/abs/physics/9807051}{\ physics/9807051}.

\bibitem{glya}
P. Glyanenko, A. Nemudry, Z.R. Ismagilov and H.J.M. Bouwmeester,
Investigation of structure, phase transition and oxygen mobility
in SrCo$_{0.8-{\rm x}}$Fe$_{0.2}$Ta$_{\rm x}$O$_{3-{\rm y}}$
mixed conductors, in preparation.

\bibitem{Kofstad}
P. Kofstad, {\em High temperature corrosion}, Elsevier, London,
1988, 558p.
\end{thebibliography}
\end{document}